\documentclass[12pt]{article}
\usepackage{geometry}
\usepackage{graphicx}
\usepackage{epsf}
\usepackage{amsmath}
\usepackage{amssymb}
\usepackage{cite}
\newcommand{\be}{\begin{equation}}
\newcommand{\ee}{\end{equation}}

\newcommand{\Rmnum}[1]{\expandafter\@slowromancap\romannumeral #1@}
\newcommand{\bea}{\begin{eqnarray}}
\newcommand{\eea}{\end{eqnarray}}
\begin{document}
\def\C{{\mathbb{C}}}
\def\R{{\mathbb{R}}}
\def\s{{\mathbb{S}}}
\def\T{{\mathbb{T}}}
\def\Z{{\mathbb{Z}}}
\def\I{{\mathbb{I}}}
\def\W{{\mathbb{W}}}
\def\Bbb{\mathbb}
\def\BZ{\Bbb Z}
\def\BR{\Bbb R}
\def\BI{\Bbb I}
\def\BW{\Bbb W}
\def\BM{\Bbb M}
\def\BC{\Bbb C} \def\BP{\Bbb P}
\def\CP{\BC\BP}
\begin{titlepage}
\title{Generalized Superconductors and Holographic Optics}
\author{}
\date{
Subhash Mahapatra, Prabwal Phukon, Tapobrata Sarkar
\thanks{\noindent E-mail:~ subhmaha, prabwal, tapo @iitk.ac.in}
\vskip0.4cm
{\sl Department of Physics, \\
Indian Institute of Technology,\\
Kanpur 208016, \\
India}}
\maketitle
\abstract{
\noindent
We study generalized holographic s-wave superconductors in four dimensional R-charged black hole and Lifshitz black hole backgrounds, in the probe limit. We first 
establish the superconducting nature of the boundary theories, and then study their optical properties. Numerical analysis indicates that a negative Depine-Lakhtakia
index may appear at low frequencies in the theory dual to the R-charged black hole, for certain temperature ranges, for specific values of the charge parameter. The 
corresponding cut-off values for these are numerically established in several cases. Such effects are seen to be absent in the Lifshitz background where this index is always positive.}
\end{titlepage}

\section{Introduction}

The use of the AdS/CFT correspondence \cite{malda},\cite{witten},\cite{klebanov} to understand strongly coupled dynamics of material systems has attracted a lot of attention of 
late. Several exciting directions of research have emerged, and there are indications that useful insight into condensed matter systems might be possible using the
gauge gravity duality \cite{hartnollrev},\cite{sachdev1}. In this context, one of the novel results in the last few years is the possibility of providing a
holographic description of the phenomenon of superconductivity \cite{Hartnoll}. While the analysis of \cite{Gubser} provided evidence of a spontaneous breaking of
gauge invariance in an Abelian Higgs model in AdS backgrounds, the work of \cite{Hartnoll} gave a holographic description of the quantum 
dynamics of the condensation of a charged operator on the boundary field theory, via a classical instability of the dual AdS black hole background. Indeed, the discovery of 
the fact that AdS black holes support scalar hair, and its implications in understanding strongly coupled condensed matter systems, have been the focus of intense research 
over the past few years. In particular, shortly after the initial results on holographic superconductors appeared, the authors of \cite{Franco} generalized the original holographic description 
of \cite{Hartnoll} to cases where the global gauge symmetry is broken via a Stuckelberg mechanism. These authors constructed gauge invariant Lagrangians, of
the Stuckelberg form, and demonstrated the possibility of a superconducting phase transition. Interestingly, their work points to the existence of a first order
phase transition, and a metastable region in the superconducting phase. 

Another interesting development in the application of holography has been in the context of optics. This is motivated in part to understand an exotic property of matter, namely
negative refraction \cite{sar},\cite{valesagorev}, via the AdS/CFT correspondence. Indeed, materials with such properties (called meta-materials) have been artificially
engineered, and find a variety of application in physics and engineering. The main property of these materials is that their dielectric permittivity, $\varepsilon$, and
the permeability, $\mu$, can both be negative, and in that case, the negative sign in the defining equation for the refractive index needs to be chosen. This corresponds
to a situation where the phase velocity of light waves in the medium is in a direction opposite to its energy flux. \footnote{The possibility of negative refraction in a curved
background has been extensively investigated in  \cite{mackay1}, \cite{mackay2}, \cite{mackay3}.} In \cite{policastro1}, a holographic description of meta-materials 
was given, and it was shown that strongly coupled field theories generically admit a negative refractive index in the hydrodynamic limit (see also \cite{bigazzi}). 
This was generalized for gauge theories dual to R-charged black hole backgrounds, in \cite{tapo1}. In the works of \cite{Gao}, \cite{amariti}, optical properties of holographic 
superconductors, were studied, and it was shown that in the probe limit, they did not exhibit phenomena of negative refraction, when the bulk theory is a five dimensional 
AdS black hole.\footnote{Superconducting meta-materials have attracted a lot of attention of late, see, e.g \cite{supmeta}.} However, it 
was shown in \cite{amariti} that inclusion of back reaction did allow for a negative refractive index at low frequencies. We point out here that importantly, in \cite{hydro},
it has been shown that in general, negative refraction in strongly coupled systems that admit a hydrodynamic description is actually a result of linear hydrodynamics of charged fluids, 
which are coupled to an external electromagnetic field. This result was obtained irrespective of holography in \cite{hydro}, and is the main reason why one expects negative refraction in 
general in the boundary theory, for small values of the wave vector and the frequency. Indeed, as we will see in the course of this paper, a holographic approach
substantiates these results.

It is important and interesting to understand electromagnetic properties of strongly coupled field theories, in particular in backgrounds different from the ones considered
in the literature thus far, and this is the task we undertake in this paper.  
Here, we consider optical properties of s-wave holographic superconductors in the probe limit, where the bulk theory is four dimensional.\footnote{In the normal phase, optical 
properties of four dimensional RN-AdS black holes have been reported in \cite{sin}.} In particular, we consider two systems of generalized holographic superconductors \cite{Franco}, 
with a background R-charged black hole \cite{Cvetic} and a background Lifshitz black hole \cite{kachru}, both of which might have interesting applications as
strongly coupled condensed matter systems. Here, we assume that the backgrounds are fixed, and that there is no back reaction. In this probe 
approximation, we first establish the superconducting properties of the boundary theories, 
i.e the existence of scalar hair below a certain temperature. Next, we study their optical properties, and calculate the Depine-Lakhtakia (DL) \cite{Depine} index by 
evaluating the retarded correlators at the boundary. We find that the DL index can be negative for sufficiently small frequencies, in contrast with the results reported in
\cite{Gao}, \cite{amariti} for five-dimensional examples, for generalized superconductors in the R-charged background. \footnote{We comment on the differences of our
results with those reported earlier in section 3.2.}
For this case, we establish the dependence of the DL index on the temperature and the black hole charge parameter and find preliminary evidence that the index of refraction 
may be negative only in a certain window of temperatures. No such effect is seen in the Lifshitz background, where the DL index is always positive. Admittedly, there are 
a number of caveats associated with these results, which will be discussed in sequel. We also mention at the outset that in our setup, there are no propagating gauge
degrees of freedom in the boundary theory, \footnote{Dynamical photons in $2 + 1$ dimensions can be introduced following \cite{Hartnoll1}, but for such degrees of freedom,  there 
is no clear conventional notion of negative refraction and we do  not consider this here.} and we talk about negative refraction by assuming that our boundary theory is 
weakly coupled to a dynamical photon. This can model experimental setups for metamaterials in two space dimensions. 

This paper is organized as follows. In the next section, we first describe the models under consideration, to fix our notations and conventions. We then proceed to show
that the models have superconducting behavior at the boundary. In section 3, the optical properties of the boundary theories are established. We end with some discussions
and our conclusions in section 4. 

\section{The Holographic Setup}

In this section we study the basic set up on the gravity side which gives a superconducting system in the boundary
theory. The first black hole background which we will be interested in, is the four dimensional R-charged black hole \cite{Cvetic} with metric
\begin{equation}
ds^{2}= -\textit{H}(r)^{-1/2}f(r)dt^{2}+\textit{H}(r)^{1/2}\left(f^{-1}(r)dr^{2}+r^2d\Omega^2\right)
\label{metric}
\end{equation}
where we define
\begin{eqnarray}
H(r)&=&H_{1}(r)H_{2}(r)H_{3}(r)H_{4}(r), ~~~~H_{i}(r)=1+\frac{\kappa_{i}r_{h}}{r}\nonumber\\
f(r)&=&k-\frac{M}{r}+r^2\prod_{i}{H_{i}(r)}
\end{eqnarray}
Here, $M$ and $\kappa_{i}$ are the mass and charge parameters of the black hole. $r_{h}$ is radius of the outer horizon, and $k$ can take two values, 0 and 1, 
which correspond to non-compact and compact horizons, respectively. In this paper, we will consider examples with planar horizons, i.e $k=0$, although all our results 
can be straightforwardly extended to the case $k=1$. In order to obtain a scalar condensate in the boundary dual, the authors in \cite{Hartnoll},\cite{Gubser} introduced 
a Abelian-Higgs model in AdS black hole backgrounds with the matter Lagrangian for the bulk theory given by
\begin{equation}
L_{matter} = -\frac{F^2}{4}-\frac{|D_{\mu}\tilde{\Psi}|^2}{2}-\frac{m^2|\tilde{\Psi}|^2}{2}
\label{matter}
\end{equation}
where $F=dA$, $D_{\mu}=\partial_{\mu}-i A_{\mu}$ and $\tilde{\Psi}$ is the complex scalar field with mass $m$. As we have mentioned, we will work in the probe limit in which 
the scalar and Maxwell fields do not back-react on the metric of eq. (\ref{metric}). Re-writing the charged scalar field $\tilde{\Psi}$ as $\tilde{\Psi}= \Psi  e^{i \alpha}$, 
the matter Lagrangian can be re-written as,
\begin{equation}
L_{matter} = -\frac{F^2}{4}-\frac{(\partial_{\mu}\Psi)^2}{2}-\frac{m^2\Psi^2}{2}-\frac{\Psi^2(\partial\alpha-A)^2}{2}
\end{equation}
where both $\Psi$ and the phase $\alpha$ are real. The local U(1) gauge symmetry in this theory is given by $A_{\mu}\rightarrow A_{\mu}+\partial_{\mu}\lambda$ together 
with $\alpha\rightarrow \alpha+\lambda$. As shown in \cite{Franco}, this model can be generalized as,
\begin{equation}
L_{matter} = -\frac{F^2}{4}-\frac{(\partial_{\mu}\Psi)^2}{2}-\frac{m^2\Psi^2}{2}-|\textrm{G}(\Psi)|(\partial\alpha-A)^2
\label{matter1}
\end{equation}
where $\textrm{G}$ is a function of $\Psi$. Now, by varying the action, it is straightforward to write down the scalar and the Maxwell's equation, and these are given by
\begin{equation}
\partial_{\mu}(\sqrt{-g}\partial^{\mu}\Psi)-\sqrt{-g}m^2\Psi-\sqrt{-g}(\partial\alpha-A)^2 \frac{d\textrm{G}}{d\Psi}=0
\end{equation}
\begin{equation}
\partial_\mu(\sqrt{-g}F^{\mu\nu})+2\sqrt{-g} \textrm{G}(\Psi)(\partial^{\nu}\alpha-A^{\nu})=0
\end{equation}
where we can use the gauge symmetry to fix the phase $\alpha=0$. Since we are interested in a superconductor like solution, we will consider the following ansatz 
\begin{equation}
\Psi=\Psi(r), ~~~~A=\Phi(r)dt
\end{equation}
Then, the equations of motion reduce to 
\begin{eqnarray}
&~& \Psi''(r) + \biggl(\frac{2}{r}+\frac{f'(r)}{f(r)}\biggr)\Psi'(r)-\frac{\sqrt{H(r)}m^2\Psi(r)}{f(r)}+\frac{H(r) \Phi^2(r)}{f^{2}(r)} \frac{d\textrm{G}}{d\Psi}=0 \nonumber\\
&~& \Phi''(r) + \biggl(\frac{2}{r}+\frac{\textit{H}'(r)}{2\textit{H}(r)}\biggr)\Phi'(r)-\frac{2\sqrt{\textit{H}(r)}\Phi(r)\textrm{G}(\Psi)}{f(r)}=0
\label{coupled}
\end{eqnarray}
In order to solve these coupled differential equations, suitable conditions at the horizon and at the boundary must be imposed. We will impose regularity conditions for 
$\Psi$ and $\Phi$ at the horizon where these fields behave as
\begin{equation}
\Phi(r_{h})=0 , ~~~~\Psi'(r_{h})=\frac{\sqrt{\textit{H}(r_{h})}m^2\Psi(r_{h})}{f'(r_{h})}
\end{equation}
Important for us will be the asymptotic expressions for $\Phi$ and $\Psi$ near the boundary,
\begin{equation}
\Phi=\mu-\frac{\rho}{r},~~~~\Psi=\frac{\Psi_{\lambda_-}}{r^{\lambda_{-}}} + \frac{\Psi_{\lambda_+}}{r^{\lambda_{+}}}
\end{equation}
where $\lambda_{\pm}=\frac{3\pm\sqrt{9+4m^2}}{2}$, and $\mu$, $\rho$ are the chemical potential and the charge density of the boundary theory, respectively. 
At the boundary both $\Psi_{\lambda_-}$ and $\Psi_{\lambda_+}$ are normalisable, and can act as sources (or vacuum expectation values) of the corresponding operators.
In this paper, we consider $\Psi_{\lambda_-}$ as the source and hence set it to zero as a boundary condition. 
\begin{figure}[t!]
\begin{minipage}[b]{0.5\linewidth}
\centering
\includegraphics[width=2.8in,height=2.3in]{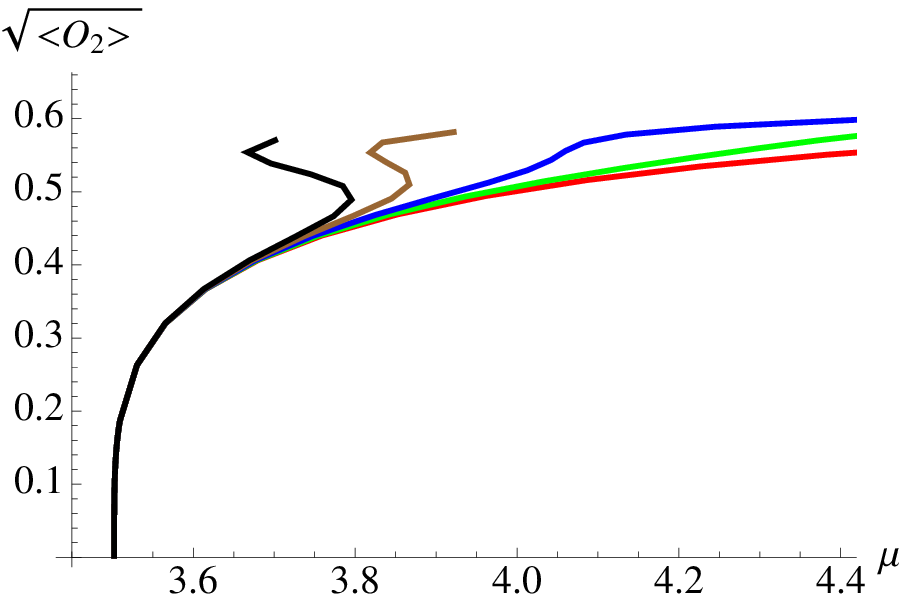}
\caption{$\sqrt{<O_{2}>}$ as a function of chemical potential for $\kappa_{1}=1, \ \kappa_{2}=\kappa_{3}=\kappa_{4}=0$. The red, green, blue, brown and black curves 
correspond to $\xi = 0, 0.2, 0.5, 1,~{\rm and}~1.5$ respectively.}
\label{O2}
\end{minipage}
\hspace{0.4cm}
\begin{minipage}[b]{0.5\linewidth}
\centering
\includegraphics[width=2.8in,height=2.3in]{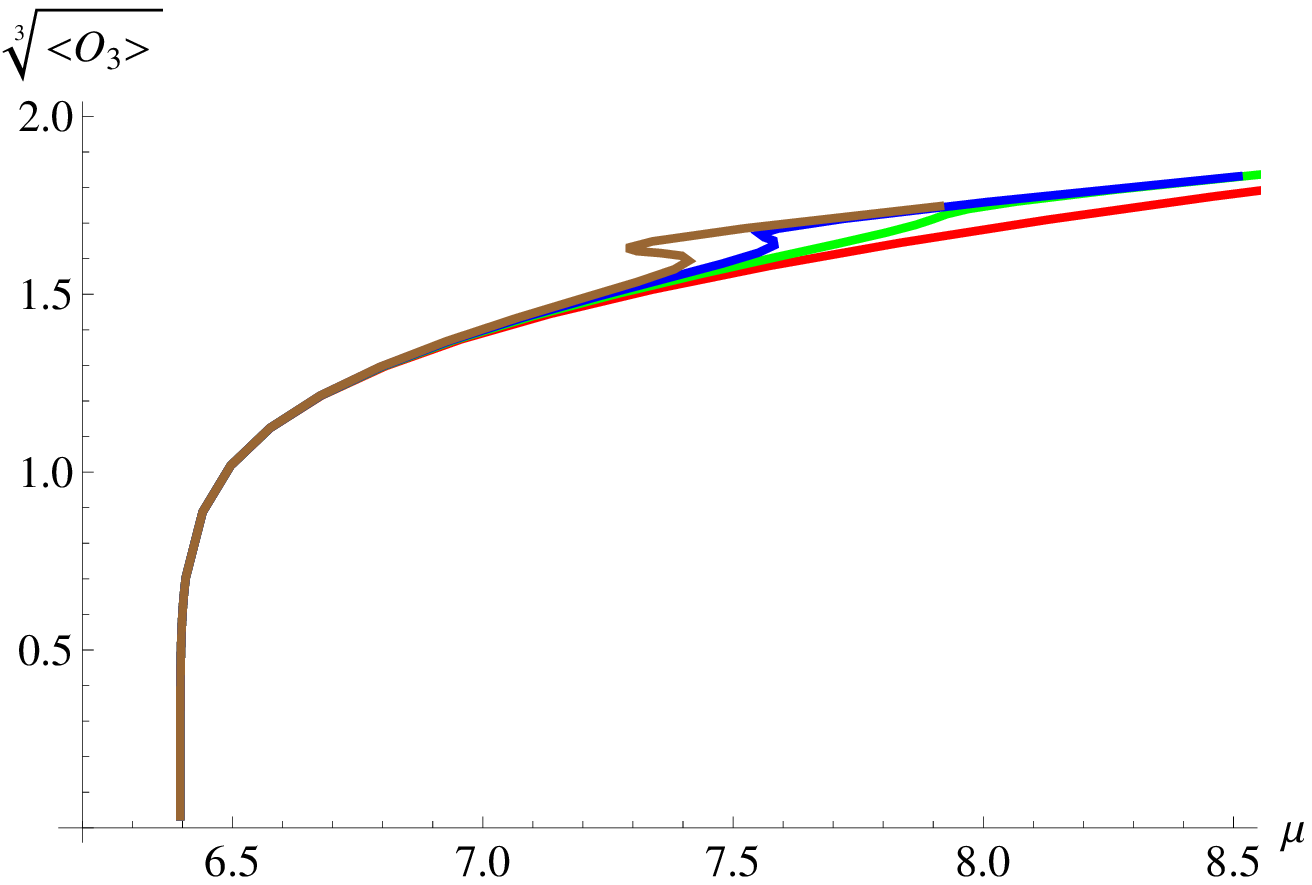}
\caption{$O_{2}$ as a function of chemical potential for the Lifshitz black hole background. The red, green, blue and brown curves corresponds to 
$\xi=0, 0.1, 0.15 \ and \ 0.2$ respectively.}
\label{O2LBH}
\end{minipage}
\end{figure}

The system in eq.(\ref{coupled}) are difficult to solve analytically, and we resort to a numerical solution.\footnote{To solve these equations numerically, we have 
used the Mathematica routine of \cite{Hartnoll}.} To simplify the numerical calculations, it is more convenient to use $z=\frac{r_h}{r}$, as is conventional in standard literature. 
In the $z$ coordinate, the horizon and the boundary are located at $z=1$ and $z=0$, respectively. For numerical calculations, we have considered $m^{2}=-2$ and the
particular form $\textrm{G}(\Psi)=\Psi^2+\xi\Psi^8$. Although $m^{2}$ is negative but it is above the Breitenlohner-Freedman bound in four dimensions, 
$m^{2}_{BF}=-9/4$ \cite{Freedman}. For this value of $m^{2}$, we get $\lambda_{\mp} = 1,2$. The condensate of the scalar operator $<O_{2}>$ in the boundary theory 
dual to the scalar field is given by
\begin{equation}
<O_{2}>\sim{\Psi_{(2)}}
\end{equation}
We need to show that our boundary theory is superconducting. In figure (\ref{O2}), we have shown the variation of condensate $<O_{2}>$ with respect to the chemical potential. 
For simplicity, here we have considered only the single R-charged case, $\kappa_{1}=1$, with all the other charges set to zero.  When the
chemical potential exceeds its critical value $\mu_{c}$, the condensate develops a nonzero vacuum expectation value which indicates the appearance of a 
superconducting phase. Below $\mu_{c}$ , the system is in the normal (or insulating) phase. We find that in our case, the critical potential is  
$\mu_{c} \sim 3.502$. This behavior indicates that our boundary theory is indeed superconducting. One should note that in the special 
case $\xi=0$, $\kappa_1 = 0$, our generalized superconductor reduces to a conventional s-wave superconductor \cite{Hartnoll}. Also, from fig.(\ref{O2}), we see that 
for sufficiently high values of $\xi$, the system exhibits a first order phase transition, similar to what was observed in \cite{cai}.

Let us now move to the second case of interest, the four dimensional Lifshitz black hole background \cite{kachru}. The conventional s-wave and p-wave superconductors in
this background have been studied in \cite{bu}. Here, we consider the metric 
\begin{equation}
ds^{2}= L^{2}\biggl(\frac{-r_{h}^{2U}}{z^{2U}}f(z)dt^2+\frac{r_{h}^2}{z^2}(dx^2+dy^2)+\frac{dz^2}{z^{2}f(z)}\biggr)
\end{equation}
where $r_h$ is the horizon radius, $z = \frac{r_h}{r}$, $L$ is the AdS length scale that we set to unity, $U$ is an anisotropy parameter, and we have $$f(z)=1-z^{(U+2)}$$
Introducing the matter lagrangian of eq.(\ref{matter1}), we obtain the equations of motion 
\begin{equation}
\Psi''(z)+\biggl(\frac{f'(z)}{f(z)}-\frac{U+1}{z}\biggr)\Psi'(z)-\frac{m^{2}\Psi(z)}{z^2f(z)}+\frac{z^{(2U-2)}\Phi^{2}(z)}{r_{+}^{2U}f^{2}(z)}\frac{d\textrm{G}(\Psi)}{d\Psi}=0
\end{equation}
\begin{equation}
\Phi''(z)+\frac{U-1}{z}\Phi'(z)-\frac{2\Phi(z)}{z^2f(z)}\textrm{G}(\Psi)=0
\end{equation}
where, as before, $\textrm{G}(\Psi)=\Psi^2+\xi\Psi^8$. Now, we impose the boundary condition near the horizon ($z=1$)  as 
\begin{equation}
\Phi(1)=0, ~~~~~\Psi'(1)=\frac{m^{2}\Psi(1)}{f'(1)}
\end{equation} 
and the asymptotic expansion of the fields near the boundary is given by  
\begin{eqnarray}
&~&{\Phi(z\rightarrow0)}  \sim \left\{ \begin{array}{ll}
\mu+\rho z^{(2-U)}~~ \mbox{for}~~ (U\neq2) \\
\rho+\mu\ln{z}~~ \mbox{for}~~ (U=2) \\ \end{array} \right. \nonumber\\
&~&\Psi(z\rightarrow0)\sim \Psi_{\Delta_{-}}z^{\Delta_{-}} + \Psi_{\Delta_{+}}z^{\Delta_{+}}
\end{eqnarray}
Here, $\Delta_{\pm}=\frac{(U+2)\pm\sqrt{(U+2)^{2}+4m^2}}{2}$. We will consider the case $m^2=-3$ and $U=2$, for which $\Delta_{\pm}=3,1$ respectively.
In fig. (\ref{O2LBH}), we have shown the condensate as a function of the chemical potential. Here, the red, green, blue and brown curves corresponds to 
$\xi=0, 0.1, 0.15 \ {\rm and} \ 0.2$ respectively. As in the R-charged example, we have considered $\Psi_{\Delta_-}$ as the source and this has been
set to zero as a boundary condition. We also mention that in this case, the critical value of the chemical potential below which the system is in the 
normal phase is $\mu_c \sim 6.395$.

Having established the superconducting property for R-charged and Lifshitz backgrounds, we will now study their optical properties. There is an obvious caveat here, namely 
the absence of a dynamical photon on the boundary. As mentioned before, we will proceed by assuming that the boundary theory is weakly coupled to a dynamical 
electromagnetic field, and that we are computing the refractive index of the system perturbatively \cite{policastro1}.

\section{Optical Properties of Generalized Superconductors}

In this section we will discuss  optical properties of generalized superconductors that we have discussed in the previous section. Here, 
we will follow the convention used in \cite{Gao}, \cite{amariti}. We first setup the system of differential equations that are used to compute 
the retarded correlators at the boundary. We then analyze these equations numerically, and establish the optical properties of the boundary system. 

\subsection{General Setup of The Problem}

The relevant quantity that is used to establish negative refractive index in a medium is called the Depine-Lakhtakia index \cite{Depine} $\eta_{DL}$, and is given by
\begin{equation}
\eta _{DL}= Re[\epsilon ]| \mu | +Re[ \mu]| \epsilon  |
\end{equation}
with negativity of the DL index indicating that the phase velocity in the medium is opposite to the direction of energy flow, i.e the system has negative refractive index. 
Computation of the DL index involves a number of steps. We begin by writing down the on shell boundary action. Subsequently, the transverse current-current correlators from the boundary 
action are obtained via the prescription of \cite{Son}. Then, we proceed by casting the momentum dependent correlators in the following form:
\begin{equation}
G_{T}\left( \omega,K \right) =G_{T}^{0}\left( \omega \right) +K^{2} G_{T}^{2}\left( \omega \right) + \cdots 
\label{GT}
\end{equation}
The permittivity and the effective permeability \cite{Landau}, can be expressed in terms of $G_{T}^{0}$ and $G_{T}^{2}$ as
\begin{equation}
\epsilon \left( \omega \right) = 1+\frac{4\pi }{\omega ^{2}} C_{em}^{2}G_{T}^{0}\left( \omega \right)
\label{eps}
\end{equation}
\begin{equation}
\mu\left( \omega \right) = \frac{1}{1- 4\pi C_{em}^{2}G_{T}^{2}\left( \omega \right) }
\label{mu}
\end{equation}
Here $C_{em}$, $\omega$ and $K$ are the EM coupling constant (set to unity for numerical computations), the frequency, and the spatial momentum, respectively.

Let us first turn to the case where the bulk theory is an R-charged black hole. We will, for computational simplicity, set $r_h = 1$.
To calculate retarded correlators \cite{Son} for the transverse currents in superconducting phase, it is enough for us to consider a perturbation of the 
Maxwell field, say $A_{x}$, on the superconducting black hole background,
\begin{equation}
A_{x}= A_{x}(r) e^{-i \omega t + iKy}
\end{equation}
In the probe limit, the equation of motion for $A_{x}$ decouples from other field equations,
\begin{equation}
A_{x}''(r) + \biggl(\frac{f'(r)}{f(r)}-\frac{H'(r)}{2 \textit{H}(r)}\biggr) A_{x}'(r) +\biggl(\frac{H(r) \omega^{2}}{f^{2}(r)}-\frac{K^{2}}{r^{2} f(r)} -
 \frac{2 \sqrt{H(r)} (\Psi^2+\xi\Psi^8)}{f(r)}\biggr) A_{x}(r) = 0
\label{eqAx}
\end{equation}
Eq.(\ref{eqAx}) must be solved with appropriate boundary conditions. We solve this equation with ingoing wave boundary condition at the horizon i.e 
\begin{equation}
A_{x} \ \alpha \ f^{\frac{-i \omega \sqrt{\textit{H}_{1}(r_{h})}}{3+2\kappa_{1}}} 
\end{equation}
As in the previous section, we have confined ourself only to the single R-Charged case, for simplicity. Also note that asymptotically, $A_{x}$ behaves as 
\begin{equation}
A_{x} = A_{x}^{(0)} + \frac{A_{x}^{(1)}}{r} + ...
\label{Axamp}
\end{equation}

From the AdS/CFT dictionary, we can identify $A_{x}^{(0)}$ and $A_{x}^{(1)}$ as the dual source and the expectation value of boundary current respectively. 
Also, the retarded correlators can be computed from the relation \cite{roberts}
\begin{equation}
G_{T} = \frac{A_x^{(1)}}{A_x^{(0)}}
\end{equation}
In order to obtain the DL index and other optical quantities in the boundary theory, we first calculate $G_{T}^0$ and $G_{T}^2$. This can be done by expanding 
$A_{x}$ in powers of $K$ in same way that $G_{T}$ is expanded in eq.(\ref{GT}),
\begin{equation}
A_{x} = A_{x0} + K^{2} A_{x2} + ...
\label{Axexp}
\end{equation}
substituting eq.(\ref{Axexp}) in the Maxwell equation of eq.(\ref{eqAx}), and separating the coefficients in powers of $K$, we obtain the differential equations for $A_{x0}$ and $A_{x2}$ as
\begin{equation}
A_{x0}''(r) + \biggl(\frac{f'(r)}{f(r)}-\frac{\textit{H}'(r)}{2 \textit{H}(r)}\biggr) A_{x0}'(r) +\biggl(\frac{\textit{H}(r) \omega^{2}}{f^{2}(r)} - \frac{2 \sqrt{\textit{H}(r)} (\Psi^2+\xi\Psi^8)}{f(r)}\biggr) A_{x0}(r) = 0
\label{r1}
\end{equation}
\begin{equation}
A_{x2}''(r) + \biggl(\frac{f'(r)}{f(r)}-\frac{\textit{H}'(r)}{2 \textit{H}(r)}\biggr) A_{x2}'(r) +\biggl(\frac{\textit{H}(r) \omega^{2}}{f^{2}(r)} - \frac{2 \sqrt{\textit{H}(r)} 
(\Psi^2+\xi\Psi^8)}{f(r)}\biggr) A_{x2}(r) - \frac{A_{x0}(r)}{r^{2} f(r)} = 0
\label{r2}
\end{equation}

The asymptotic forms of $A_{x0}$ and $A_{x2}$ can be found from eqs. (\ref{r1}) and (\ref{r2}) as : 
\begin{equation}
A_{x0} = A_{x0}^{(0)} + \frac{A_{x0}^{(1)}}{r} + ..., ~~~~~A_{x2} = A_{x2}^{(0)} + \frac{A_{x2}^{(1)}}{r} + ...
\end{equation}
and therefore, we obtain
\begin{equation}
G_{T}^0 = \frac{A_{x0}^{(1)}}{A_{x0}^{(0)}}, \ \ \ \
G_{T}^2 = \frac{A_{x0}^{(1)}}{A_{x0}^{(0)}} \biggl(\frac{A_{x2}^{(1)}}{A_{x0}^{(1)}} - \frac{A_{x2}^{(0)}}{A_{x0}^{(0)}} \biggr)
\label{GT0}
\end{equation}
We substitute eq.(\ref{GT0}) into eq.(\ref{eps}), (\ref{mu}) and obtain $\varepsilon(\omega)$, $\mu(\omega)$ and $n_{DL}$ in terms of $A_{x0}^{(0)}$, $A_{x0}^{(1)}$, $A_{x2}^{(0)}$
and $A_{x2}^{(1)}$. We solve these equations numerically and as before, use the $z=\frac{r_h}{r}$ to simplify the numerics.

An entirely similar analysis follows for the Lifshitz black hole background. In what follows, we will, for algebraic simplicity, use the coordinate $z = \frac{r_h}{r}$. 
With the gauge field perturbation $A_{x}=A_{x}(z)e^{(-i\omega t + iKy)}$, we obtain the equation of motion (we set $r_h = 1$)
\begin{equation}
A_{x}''(z)+\biggl(\frac{f'(z)}{f(z)}-\frac{(U-1)}{z}\biggr)A_{x}'(z)+\biggl(\frac{\omega^{2}z^{2U-2}}{r_{h}^{2U}f^{2}(z)} - \frac{K^2}{r_{h}^{2U}f(z)} - \frac{2\textrm{G}(\Psi)}{z^{2}f(z)}\biggr)A_{x}(z)=0
\end{equation}
following a similar procedure as in the R-charged example, we obtain 
\begin{equation}
A_{x0}''(z)+\biggl(\frac{f'(z)}{f(z)}-\frac{(U-1)}{z}\biggr)A_{x0}'(z)+\biggl(\frac{\widetilde{\omega}^{2}z^{2U-2}}{f^{2}(z)} - \frac{2\textrm{G}(\Psi)}{z^{2}f(z)}\biggr)A_{x0}(z)=0
\end{equation}
\begin{equation}
A_{x2}''(z)+\biggl(\frac{f'(z)}{f(z)}-\frac{(U-1)}{z}\biggr)A_{x2}'(z)+\biggl(\frac{\widetilde{\omega}^{2}z^{2U-2}}{f^{2}(z)} - \frac{2\textrm{G}(\Psi)}{z^{2}f(z)}\biggr)A_{x2}(z) -\frac{A_{x0}}{f(z)} =0
\end{equation}
where $\widetilde{\omega}=\omega/r_{+}^{U}$. Now, we use the asymptotic forms 
\begin{equation}
A_{x0}(z) =  A_{x0}^{(0)} + z^2A_{x0}^{(2)} + \cdots,~~~~~ A_{x2}(z) =  A_{x2}^{(0)} + z^2A_{x2}^{(2)} + \cdots
\end{equation}
and after a little bit of algebra, we get
\begin{equation}
G_{T}^0= \frac{2A_{x0}^{(2)}}{A_{x0}^{(0)}},~~~~~~G_{T}^2= \frac{2A_{x0}^{(2)}}{A_{x0}^{(0)}}\biggl(\frac{A_{x2}^{(2)}}{A_{x0}^{(2)}} -
\frac{A_{x2}^{(0)}}{A_{x0}^{(0)}} \biggr)
\end{equation}
A comment is in order here. Typically, in five dimensional setups, one encounters presence of divergences that need to be removed by renormalization procedures.
This leaves a free coefficient that is fixed from the requirement that for large frequencies, the system effectively behaves like the vacuum, i.e the electric permittivity 
tends to one in this limit. In our four dimensional examples, such ambiguities do not exist. Indeed, as we will show in sequel (see figs. (\ref{epsRe}) and (\ref{epsIm})), the permittivity
approaches unity for large frequencies, as expected. 

\subsection{Numerical Analysis and Results}

\begin{figure}[t!]
\begin{minipage}[b]{0.5\linewidth}
\centering
\includegraphics[width=2.8in,height=2.3in]{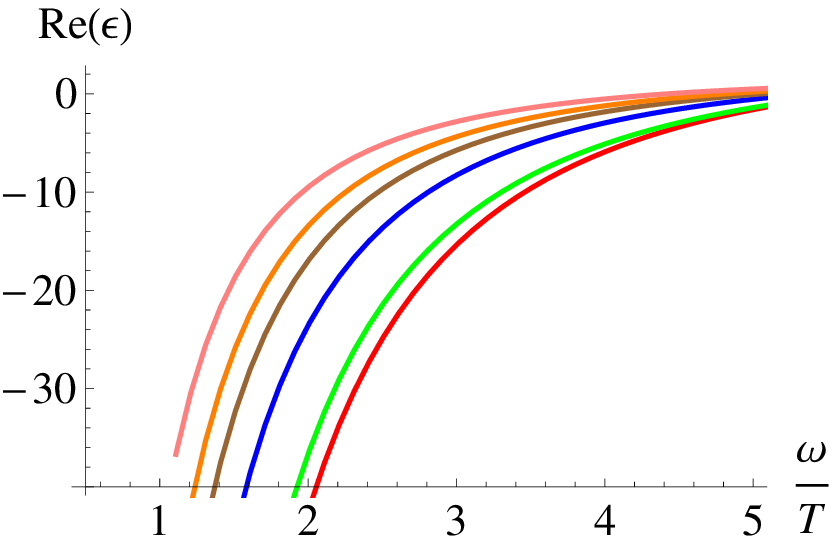}
\caption{$Re(\varepsilon)$ as a function of $\omega/T$ for $\xi=0.2$ at $T = 0.81 T_{c}$.}
\label{epsRe}
\end{minipage}
\hspace{0.4cm}
\begin{minipage}[b]{0.5\linewidth}
\centering
\includegraphics[width=2.8in,height=2.3in]{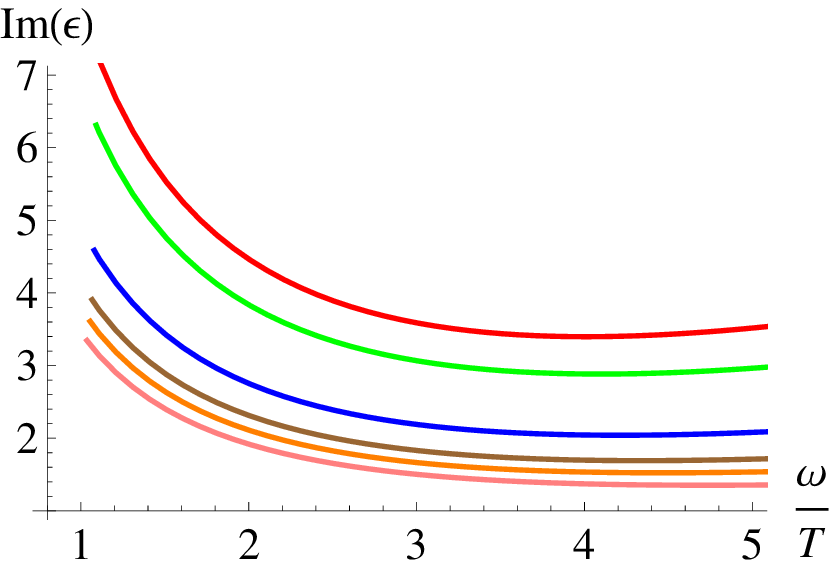}
\caption{$Im(\varepsilon)$ as a function of $\omega/T$ for $\xi=0.2$ at $T = 0.81 T_{c}$.}
\label{epsIm}
\end{minipage}
\end{figure}

\begin{figure}[h!]
\begin{minipage}[b]{0.5\linewidth}
\centering
\includegraphics[width=2.8in,height=2.3in]{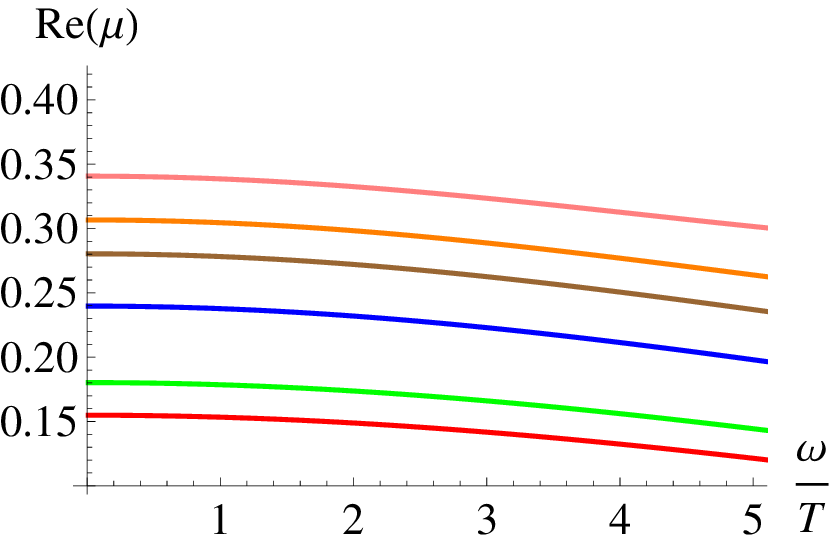}
\caption{$Re(\mu)$ as a function of $\omega/T$ for $\xi=0.2$ at $T = 0.81 T_{c}$.}
\label{muRe}
\end{minipage}
\hspace{0.4cm}
\begin{minipage}[b]{0.5\linewidth}
\centering
\includegraphics[width=2.8in,height=2.3in]{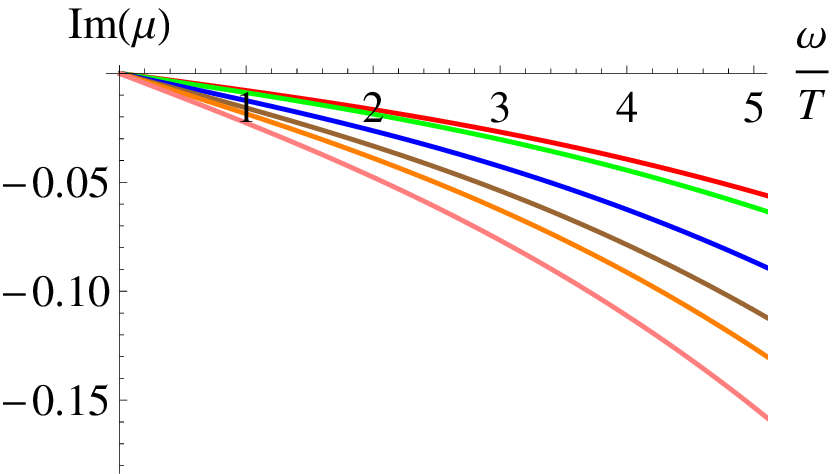}
\caption{$Im(\mu)$ as a function of $\omega/T$ for $\xi=0.2$ at $T = 0.81 T_{c}$.}
\label{muIm}
\end{minipage}
\end{figure}

In this subsection we will numerically analyze the optical properties of the boundary superconducting theory. \footnote{In what follows, the temperatures are
measured in appropriate units of $\rho$, the charge density of the boundary theory.} As before, we will consider the single R-charged case. 
In figs.(\ref{epsRe}), (\ref{epsIm}), (\ref{muRe}) and (\ref{muIm}), we plot ${\rm Re}(\varepsilon)$,
${\rm Im} (\varepsilon)$, ${\rm Re} (\mu)$, and ${\rm Im} (\mu)$ respectively, as a function of $\omega/T$. 
Here, we have chosen $T= 0.81 T_c$ (so that our system is in the superconducting phase) and $\xi = 0.2 $. 
In figs.(\ref{epsRe})-(\ref{muIm}) the red, green, blue, brown, orange and pink curves
 corresponds to $\kappa_{1} = 0$, $1$, $5$, $10$, $15$ and $25 $ respectively.\footnote{In these and all subsequent graphs, the same 
color coding will be used. Namely, the red, green, blue, brown, orange and pink curves correspond to $\kappa_{1}$= $0$, $1$, $5$, $10$, $15$ and $25$ respectively.
We do not mention this in sequel.} 
Our results for the permittivity and the effective permeability are qualitatively similar to those obtained in \cite{Gao}, \cite{amariti}. 
As expected, ${\rm Re} (\varepsilon)$ is negative at low frequencies and ${\rm Im} (\varepsilon)$ is always positive and has a pole at zero frequency for all $\kappa_{1}$. 
The issue about the permeability is however more subtle. As we see from fig.(\ref{muIm}), the imaginary part of the permeability is negative. In fact, this is the case
in all probe limit analyses that have appeared in the literature, and as pointed out in \cite{amariti}, this might be problematic, although we have  used an effective
permeability that is not an observable. We are unable to comment on this further, but point out here that the issue of sign of the imaginary part of the permeability
is still not a settled issue in the optics literature (see, e.g \cite{negimmu}, although this work deals with real permeability). 

\begin{figure}[t!]
\begin{minipage}[b]{0.5\linewidth}
\centering
\includegraphics[width=2.7in,height=2.3in]{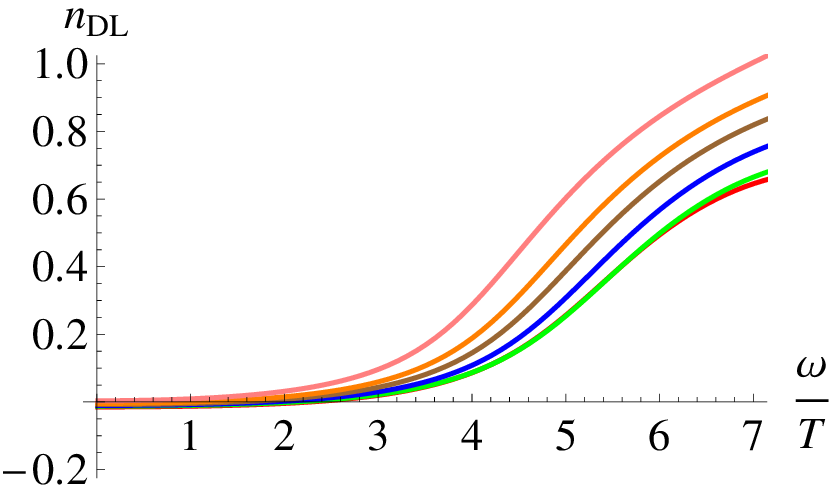}
\caption{$n_{DL}$ as a function of $\omega/T$ for $\xi=0.2$ at $T = 0.81 T_{c}$.}
\label{NDLzeta0.2temp0.81Tc}
\end{minipage}
\hspace{0.4cm}
\begin{minipage}[b]{0.5\linewidth}
\centering
\includegraphics[width=2.7in,height=2.3in]{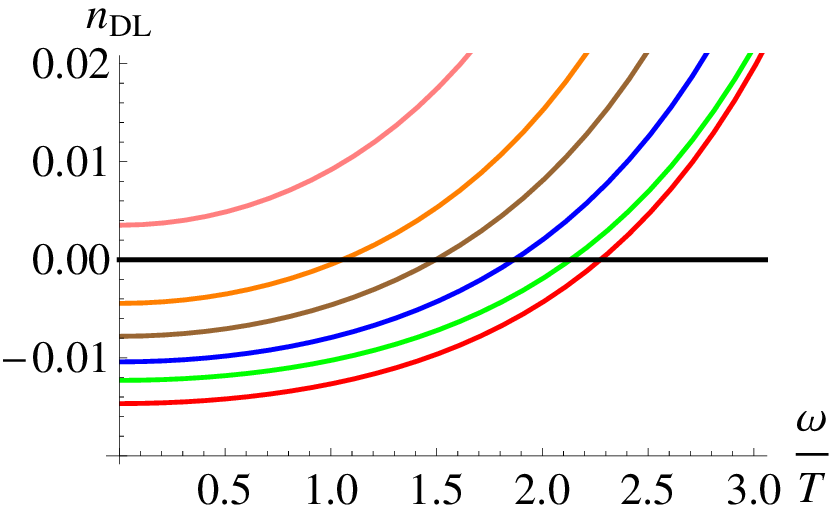}
\caption{Low frequency behavior of $n_{DL}$ for $\xi=0.2$ at $T = 0.81 T_{c}$.}
\label{lowNDLzeta0.2temp0.81Tc}
\end{minipage}
\end{figure}

Our main purpose here is to calculate $n_{DL}$ and this is shown in figure (\ref{NDLzeta0.2temp0.81Tc}). 
At high frequencies, $n_{DL}$ has the same qualitative features found in \cite{Gao}, \cite{amariti}. But at low frequencies, a significant difference emerges, with the 
appearance of negative $n_{DL}$. To illustrate the point, in fig.(\ref{lowNDLzeta0.2temp0.81Tc}), we have plotted $n_{DL}$ for low 
frequencies and one can see the emergence of negative DL index, below a certain value of $\omega/T$. 

Let us elaborate on this a bit further. For the moment, we focus on the
simpler situation, $\xi = 0$. Here, our numerical analysis indicates that there is no negative DL index for $T = 0.85 T_c$, for any value of $\kappa_1$. For $T = 0.83T_c$, we 
find that negative refraction is allowed for small values of $\kappa_1$, upto $\kappa_1 \sim 10$. When the temperature is further reduced to $T = 0.75T_c$, we find 
the possibility of a negative DL index upto $\kappa_1 \sim 50$. Now, at low values of the temperature, back-reaction effects might be important, and numerical analysis in
the probe limit may not be fully trusted. However, we find an indication that there might be a window of temperatures for which negative refraction is allowed. 
For example, setting $\xi = 0$, $\kappa_1 = 0$, negativity in the DL index sets in as the temperature is lowered from its critical value to 
$\sim 0.83T_c$. At $T = 0.1T_c$, however, this is not observed. We emphasize that this is simply an indication, and needs to be investigated further, 
with inclusion of back-reaction effects. 
\begin{figure}[t!]
\begin{minipage}[b]{0.5\linewidth}
\centering
\includegraphics[width=2.7in,height=2.3in]{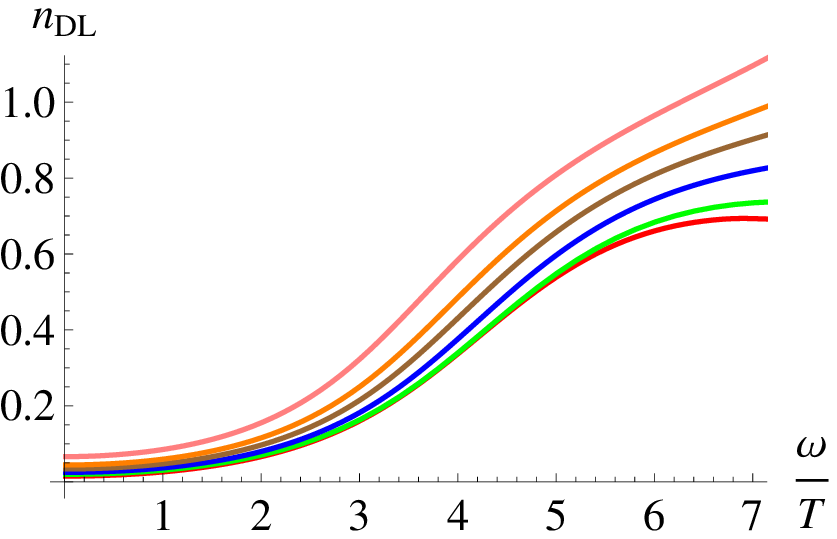}
\caption{$n_{DL}$ a function of $\omega/T$ for $\xi=0.2$ at $T = 0.86 T_{c}$.}
\label{NDLzeta0.2temp0.86Tc}
\end{minipage}
\hspace{0.4cm}
\begin{minipage}[b]{0.5\linewidth}
\centering
\includegraphics[width=2.7in,height=2.3in]{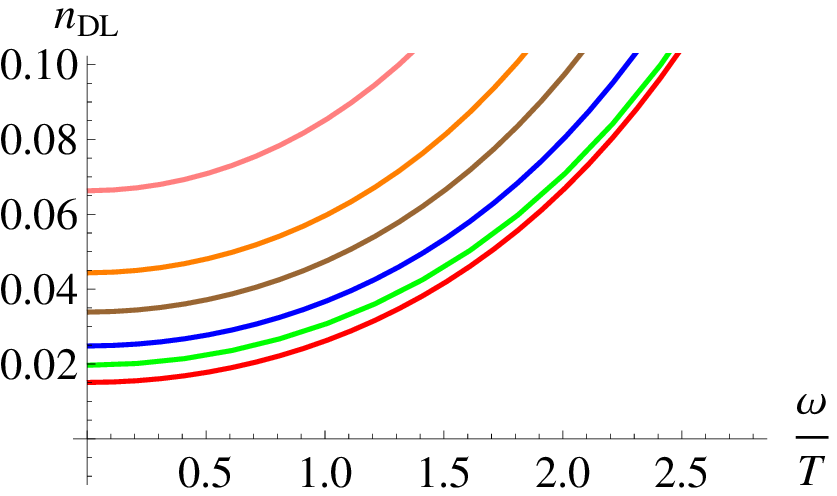}
\caption{Low frequency behavior of $n_{DL}$ for $\xi=0.2$ at $T = 0.86 T_{c}$. }
\label{lowNDLzeta0.2temp0.86Tc}
\end{minipage}
\end{figure}
Further, our results indicate that, at a particular $\xi$ and a fixed temperature, there exists a critical $\kappa_{1c}$ above which a negative DL index cannot occur, 
irrespective of how low the frequency is. This can be seen, for example, from fig.(\ref{lowNDLzeta0.2temp0.81Tc}), where no negative DL index occurs above a particular
value of $\kappa$. A more useful statement is that for 
a fixed value of $\xi$ and $\kappa_1$, the system might go from a positive refraction ``phase'' to a negative refraction ``phase'' by varying the temperature.
To justify this, we have plotted $n_{DL}$ in 
figs.(\ref{NDLzeta0.2temp0.86Tc}) and (\ref{lowNDLzeta0.2temp0.86Tc}) at a 
temperature $T = 0.86 T_{c}$. It is observed that negative refraction which occurs for $\kappa_{1}=0$, $1$, $5$, $10$ and $15$, at $T=0.81T_{c}$,
(fig.(\ref{lowNDLzeta0.2temp0.81Tc})), disappears at $T=0.86T_{c}$. Interestingly, qualitatively similar results have
been the topic of discussion in the optics community of late, although in a somewhat different context (see, e.g. \cite{temp1}), but we make no claims beyond a naive similarity 
of those results with ours. 
\begin{figure}[t!]
\begin{minipage}[b]{0.5\linewidth}
\centering
\includegraphics[width=2.7in,height=2.3in]{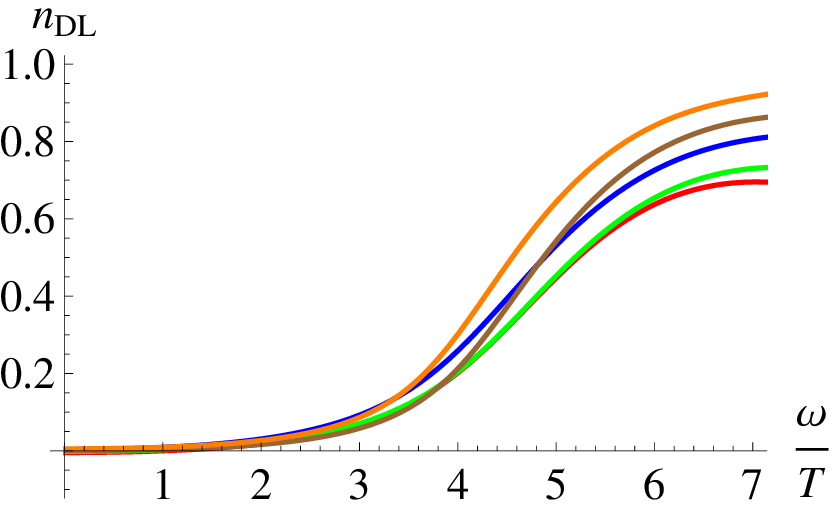}
\caption{$n_{DL}$ as a function of $\omega/T_{c}$ for $\xi=1$ at $T = 0.86 T_{c}$.}
\label{NDLzeta1temp0.86Tc}
\end{minipage}
\hspace{0.4cm}
\begin{minipage}[b]{0.5\linewidth}
\centering
\includegraphics[width=2.7in,height=2.3in]{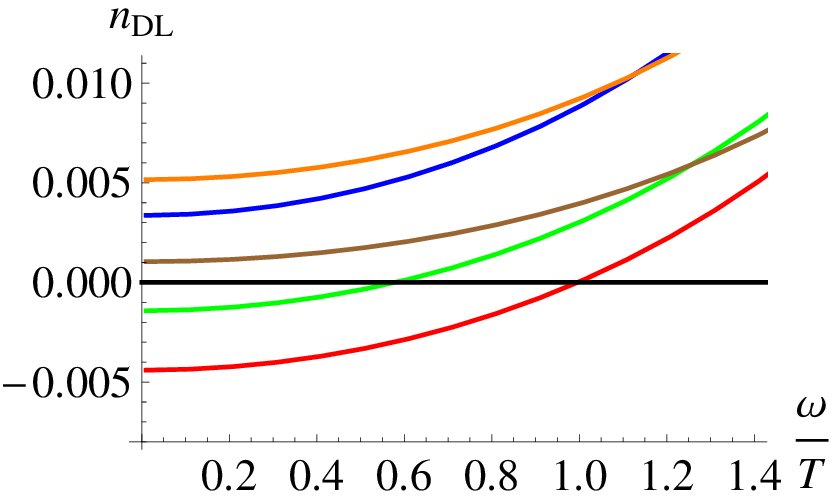}
\caption{Low frequency behavior of $n_{DL}$ for $\xi=1$ at $T = 0.86 T_{c}$.}
\label{lowNDLzeta1temp0.86Tc}
\end{minipage}
\end{figure}

We have also calculated $n_{DL}$ for other values of $\xi$. In figs.(\ref{NDLzeta1temp0.86Tc}) and (\ref{lowNDLzeta1temp0.86Tc}), we have used $\xi=1$ 
and plotted $n_{DL}$ at $T=0.86T_{c}$. Expectedly, the essential features of our analysis are similar with the $\xi=0.2$ case. 
In figure (\ref{lowNDLzeta1temp0.86Tc}), we get a negative DL index at low frequencies for $\kappa_{1}=0$ and $1$. By comparing fig.(\ref{lowNDLzeta0.2temp0.86Tc}) 
with fig.(\ref{lowNDLzeta1temp0.86Tc}), we see that the possibility for negative refraction for fixed $\kappa_{1}$,  increases with $\xi$. 

It is important to point out the differences of our results with the ones reported in five dimensional AdS backgrounds, as mentioned in the introduction. Specifically, near
the critical point, $T\simeq T_{c}$, when the condensate is small, one can, for ${\rm AdS}_5$ backgrounds, perform analytic calculations to obtain approximate anlytic expressions 
for the transverse current correlator. The main idea of \cite{herzoganalytic}, \cite{hydro} is to expand the bulk fields in terms of certain small parameters wherein one needs
to solve for the Heun equations. In four dimensional backgrounds as considered in this paper, this procedure fails to produce analytic results. However we point out
that these analytic results will be valid only close to $T_c$ and for our case, we do find that the DL index is positive close to $T \simeq T_c$, similar to the five dimensional 
examples. It is only when we go below such values of the temperature that negative refraction appears, which we believe is associated with the nature of the four dimensional solution. 
For the Lifshitz background, we find that the methods of  \cite{herzoganalytic}, \cite{hydro} yield complicated analytic expressions for the current correlator near the critical temperature
in terms of hypergeometric functions, which do not yield useful results. 

A word about the range of validity of our results is in order. Usually, the retarded current-current correlator $G^R_{xx}$ is written down in a
compact form :
\begin{equation}
G^R_{xx}=G_T^0(\omega) + K^2 G_T^2(\omega)+ \cdots
\end{equation}
The series is restricted to the first two terms, provided,
\begin{equation}
|\frac{G_T^2(\omega)K^2}{G_T^0(\omega)} |\ll 1 
\end{equation} 
Or equivalently, one can write
\begin{equation}
|\frac{G_T^2(\omega)n^2}{G_T^0(\omega)} |\omega^2\ll 1 
\label{validity1}
\end{equation}
Consequently, the $\epsilon-\mu$ analysis is valid only for those frequencies where the above constraint is not violated.
\begin{figure}[t!]
\begin{minipage}[b]{0.5\linewidth}
\centering
\includegraphics[width=2.7in,height=2.3in]{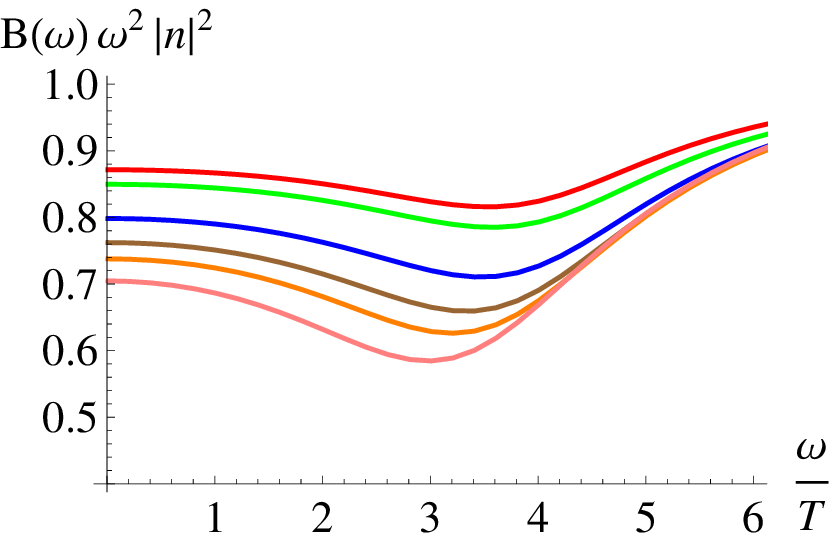}
\caption{$B(\omega)\omega^{2}|n|^{2}$ as a function of $\omega/T$ for $\xi=0.2$ at $T = 0.86 T_{c}$.}
\label{valid.86}
\end{minipage}
\hspace{0.4cm}
\begin{minipage}[b]{0.5\linewidth}
\centering
\includegraphics[width=2.7in,height=2.3in]{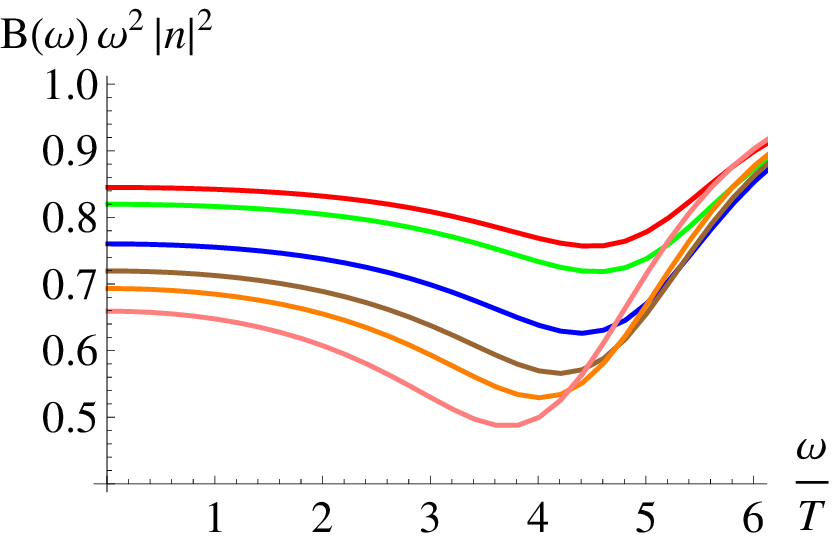}
\caption{$B(\omega)\omega^{2}|n|^{2}$ as a function of $\omega/T$ for $\xi=0.2$ at $T = 0.81 T_{c}$.}
\label{valid.81}
\end{minipage}
\end{figure} 
\begin{figure}[t!]
\begin{minipage}[b]{0.5\linewidth}
\centering
\includegraphics[width=2.7in,height=2.3in]{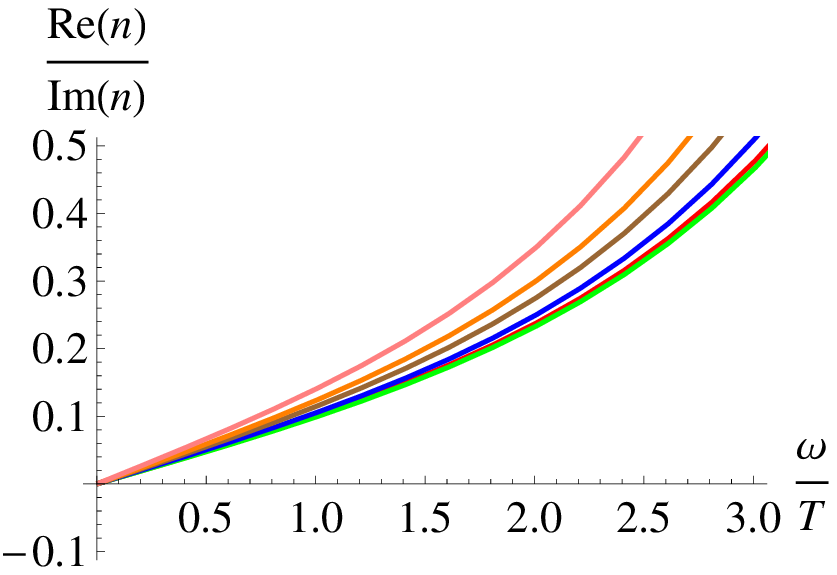}
\caption{$Re(n)/Im(n)$ as a function of $\omega/T$ for $\xi=0.2$ at $T = 0.86 T_{c}$.}
\label{diss1}
\end{minipage}
\hspace{0.4cm}
\begin{minipage}[b]{0.5\linewidth}
\centering
\includegraphics[width=2.7in,height=2.3in]{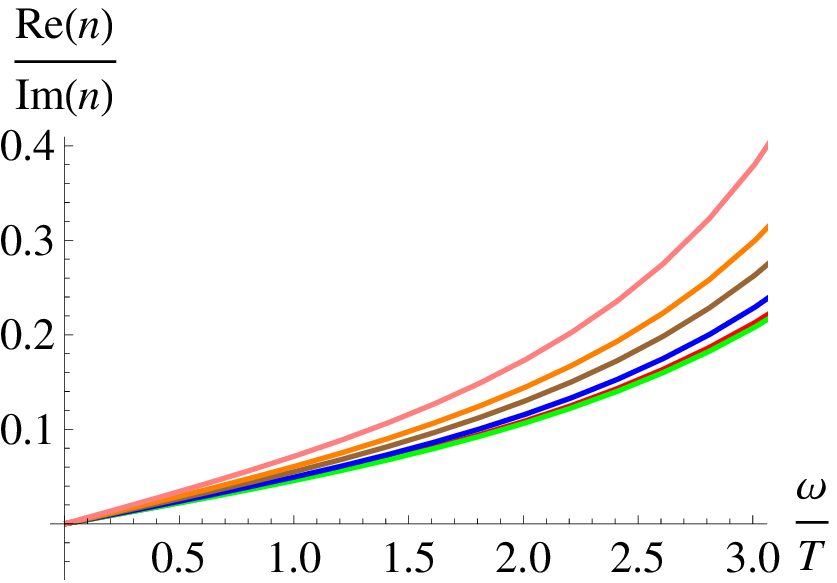}
\caption{$Re(n)/Im(n)$ as a function of $\omega/T$ for $\xi=0.2$ at $T = 0.81 T_{c}$.}
\label{diss2}
\end{minipage}
\end{figure}

In our case, we plot $|\frac{G_T^2(\omega)n^2}{G_T^0(\omega)} |\omega^2\ $  in figs.(\ref{valid.86}) and (\ref{valid.81}), where in the figures we have
called $|\frac{G_T^2(\omega)}{G_T^0(\omega)}| = B$. One can see that, within the plotted frequency range,
$|\frac{G_T^2(\omega)n^2}{G_T^0(\omega)} |\omega^2$ is less than unity for ranges of frequencies in which the DL index is negative. Here we encounter a further
caveat in our results. Namely, that eq.(\ref{validity1}) is not very strictly satisfied. This is a problem with the probe approximation, and could be related to the 
appearance of a negative imaginary part of the magnetic permeability that we have encountered before. Currently we do not have
a resolution to this, and it will be important to understand this issue further, including back reaction effects. 

We also need to study the dissipation effects, which is obtained from the ratio $\frac{Re[n]}{Im[n]}$. This is illustrated in figs.(\ref{diss1}) and (\ref{diss2}). 
The propagation to dissipation ratio unfortunately tends to zero at very low frequencies, implying
that in part of the domain of negative refraction, propagation of electromagnetic waves is virtually absent. However, we remind the reader that this is a generic
result in any holographic setup, and is not unexpected. Back reaction effects might make the ratio somewhat larger, 
as pointed out in \cite{amariti}. Further, we note here that unlike real metamaterials, ${\rm Re}(n)$ and ${\rm Im}(n)$ are of the same sign. 
This problem also is likely to go away if backreaction effects are included.  

In this paper we have only presented the analysis for $n_{DL}$ and other optical quantities for the single R-charged cases, but the same analysis can be generalized to multiple R-charged 
backgrounds. Preliminary results involving multiple charge examples indicate that the essential features of $n_{DL}$ remains the same as in the single charge case.
\begin{figure}[t!]
\begin{minipage}[b]{0.5\linewidth}
\centering
\includegraphics[width=2.7in,height=2.3in]{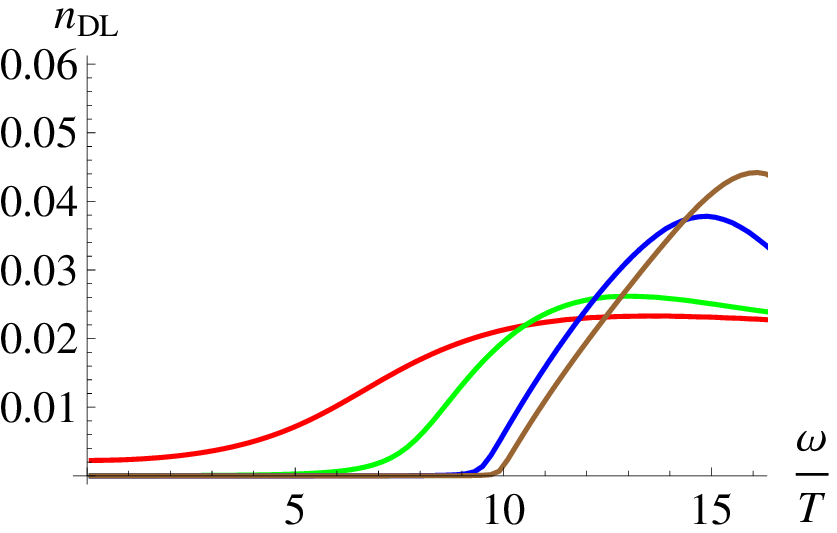}
\caption{$n_{DL}$ as a function of $\omega/T$ at $T = 0.80 T_{c}$ for the Lifshitz black hole. The red, green, blue and brown curves corresponds to $\xi=0, 0.1, 0.15 \ and \ 0.2$ respectively. 
It is seen that there is no negative refraction.}
\label{lif1}
\end{minipage}
\hspace{0.4cm}
\begin{minipage}[b]{0.5\linewidth}
\centering
\includegraphics[width=2.7in,height=2.3in]{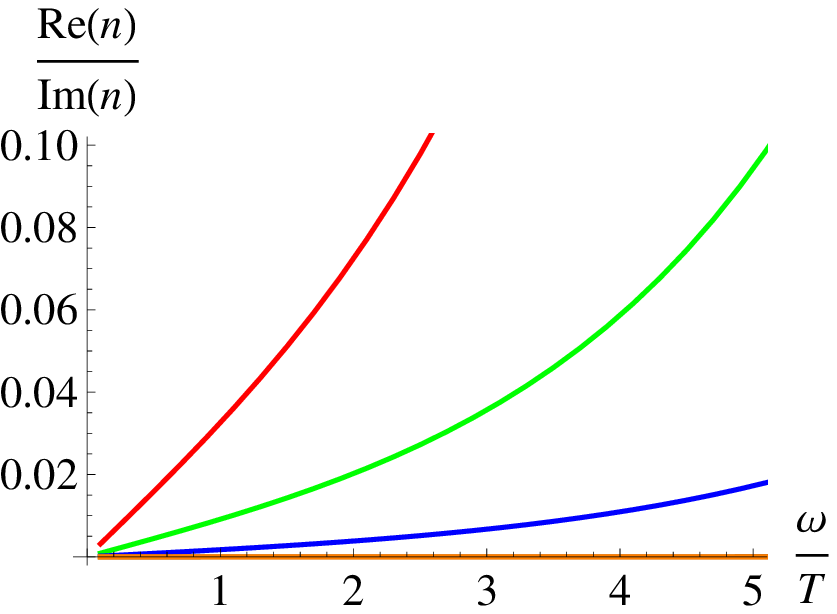}
\caption{${\rm Re}(n)/{\rm Im}(n)$ for the Lifshitz black hole, as a function of $\omega/T$ for $\xi=0$. The red, green, blue, brown and orange curves corresponds to 
$T/T_{c}=0.80, 0.65, 0.50, 0.20 \ and \ 0.10$ respectively. The last two lie almost along the $\omega/T$ axis.}
\label{lif2}
\end{minipage}
\end{figure}
For the Lifshitz background, we do not find any evidence of a negative DL index in the probe limit. We do not discuss the results in details, but present them in 
figs.(\ref{lif1}) and (\ref{lif2}), showing the DL index and the propagaion to dissipation ratio, respectively. 

\section{Conclusions}

In this paper, we have studied optical properties of generalized holographic superconductors corresponding to four dimensional R-charged black hole and Lifshitz black hole 
backgrounds, in the probe limit. As a special case, our R-charged example reduces to the four dimensional SAdS background, whose five dimensional form was studied in the same 
limit in \cite{Gao}.

In this paper, we have seen that for the R-charged backgrounds, at small enough frequencies and below a cut-off value of the charge parameter $\kappa_{1c}$, numerical 
analysis indicates that the superconducting phase can exhibit a negative DL index. By increasing the temperature, this behavior disappears indicating that 
there might be a window of temperatures for which the system exhibits negative refraction. We have analyzed in details the optical properties of the system dual 
to R-charged black holes by varying the control parameters of the theory as well as the temperature. Qualitatively, we get predictions similar to \cite{temp1}, and it will be 
interesting to analyze this issue further, to see if one can get closer to realistic systems. For the Lifshitz black hole background, there is no possibility of a negative DL index. 
Whether this sheds some light on optical properties of strange metals might be an interesting topic to study. 

Our results should be contrasted with those of \cite{Gao}, \cite{amariti}, where a negative DL index was not observed in the probe limit, for five dimensional 
black hole backgrounds. In particular, it was shown \cite{amariti} that negative refraction occurs only when the matter fields backreact on the bulk metric. 
The appearance of a negative DL index in our R-charged example (modulo the various caveats that we have discussed) is possibly due to the different nature of the boundary 
theory here. Also, as we have seen, the tunable R-charge parameter plays an important role in determining the optical properties of the boundary theory. 

Our analysis is limited by the fact that we worked in the probe limit, and admittedly, this led to several caveats. It will be interesting to study the effect of holographic 
superconductors in R-charged backgrounds and in Lifshitz backgrounds by including effects of back reaction, although the numerics might be substantially more 
complicated. We leave such an analysis for the future. 

\begin{center}
{\bf Acknowledgements}
\end{center}
It is a pleasure to thank A. Amariti and D. Forcella for very useful comments. The work of SM is supported by grant no. 09/092(0792)-2011-EMR-1 from CSIR, India.\\

\end{document}